# On-chip 1 TOPS Hyperdimensional Photonic Tensor Core using a WDM Silicon Photonic Coherent Crossbar

S. Kovaios, I. Roumpos, A. Tsakyridis, M. Moralis-Pegios,
D. Lazovsky, K. Vyrsokinos and N. Pleros

***Abstract*— We demonstrate an on-chip 0.96 TOPS hyperdimensional photonic tensor core by utilizing a time-space-wavelength multiplexed silicon photonic Crossbar (Xbar). The novel architecture relies on serializing the large matrix-vector or tensor-vector products by unfolding multiply and accumulation operations over time domain, while simultaneously distributing the computational workload over different spatial and wavelength channels. We experimentally demonstrate the operation of a 4-channel 2-input TSWDM Xbar that incorporates 56 GHz electro-absorption modulators (EAMs) and 4-channel integrated multiplexing stages. Its successful operation as a 4×2×1 tensor-vector multiplication unit demonstrated an average error of 3.9%. Its performance as a photonic AI accelerator was also evaluated in the classification task of the Iris dataset, presenting experimental accuracies of 93.3% at data rates between 4×10 and 4×30 GBd, reaching 83.3% when the data rate increases to 4×60 GBd. Finally, we discuss the TSWDM Xbar scalability potential, revealing that the inclusion of a WDM scheme in the SDM architecture reduces the operating laser power, feasibly boosting the potential of constructing photonic accelerators with computational throughput in the POPS regime.**

## I. Introduction

The burst of artificial intelligence (AI) applications has turned the demand for highly efficient computing systems into an urgent necessity [1]. This has stimulated a huge effort into the development of specialized units for linear algebra operations, such as matrix-vector (MVM) and tensor-vector (TVM) multiplications, where achieving optimal energy efficiencies and high computational throughput credentials remains a highly challenging goal. Photonic accelerators appear as highly promising candidates for next generation systems, since they combine high bandwidth and high energy efficiency credentials [2]-[5]. Their potential for leading the race in energy efficient AI computing has been confirmed so far by the numerous experimental demonstrations that have been reported during the last years [8][7]-[29] and penetrated the sub-pJ/MAC computational efficiency regime [6], implying their potential to reach in principle even aJ/MAC capabilities [7].

However, the impressive architectural and technological advances in the field of Photonic Neural Networks (PNNs) [6] are still facing severe challenges in meeting the trainable parameter requirements of today's AI models and at the same time deliver a total computational power in the beyond 100's TOPS regime. SDM layouts have been primarily implemented through coherent photonic meshes [8]-[10], presenting, however, a rather limited scalability potential since insertion losses increase with their fan-in/fan-out dimensions. On the other hand, WDM schemes that exploit the excess bandwidth available in photonic systems to parallelize computational tasks have been widely adapted [11]-[13] with SDM layouts, even though the generation of a substantial number of channels still poses a challenge for next generation systems [14]. Finally, TDM techniques that rely on unfolding mathematical operations in the time domain [15]-[18] have marginalized the necessity of matching hardware dimensions with large NN's parameter space, introducing however additional latency and

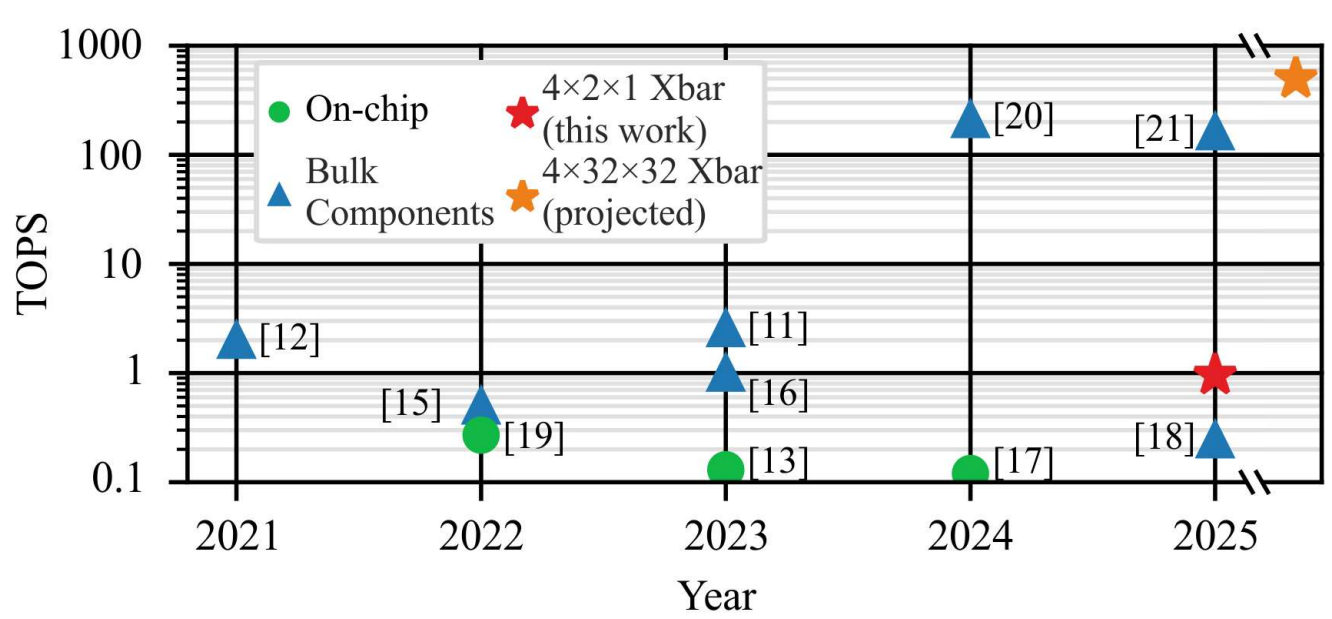

Fig. 1. Computational throughput of photonic accelerators

Manuscript received XX XX, 2025; revised Month XX, 2025; accepted Month XX, 2025. Date of publication Month XX, 2025; date of current version Month XX, 2025. This work was supported by the European Commission through the HORIZON Projects SIPHO-G (101017194), PARALIA (101093013), GATEPOST (101120938) and HAETAE (101194393). (*corresponding author:* sdkovaios@csd.auth.gr)

S. Kovaios, A. Tsakyridis, M. Moralis-Pegios and N. Pleros are with the School of Informatics, Aristotle University of Thessaloniki, 54124, Thessaloniki, Greece and the Center for Interdisciplinary Research & Innovation (CIRI-AUTH), Balkan Center, 57001, Greece (e-mail: sdkovaios@csd.auth.gr; atsakyrid@csd.auth.gr; mmoralis@csd.auth.gr; npleros@csd.auth.gr).

I. Roumpos and K. Vyrsokinos are with the School of Physics, Aristotle University of Thessaloniki, 54124, Thessaloniki, Greece and the Center of Interdisciplinary Research and Innovation (CIRI-AUTH), Balkan Center, 57001, Greece (e-mail: ioroumpo@auth.gr; kv@auth.gr).

D. Lazovsky is with Celestial AI, Campbell, CA 95008, USA. (e-mail: dl@celestial.ai)

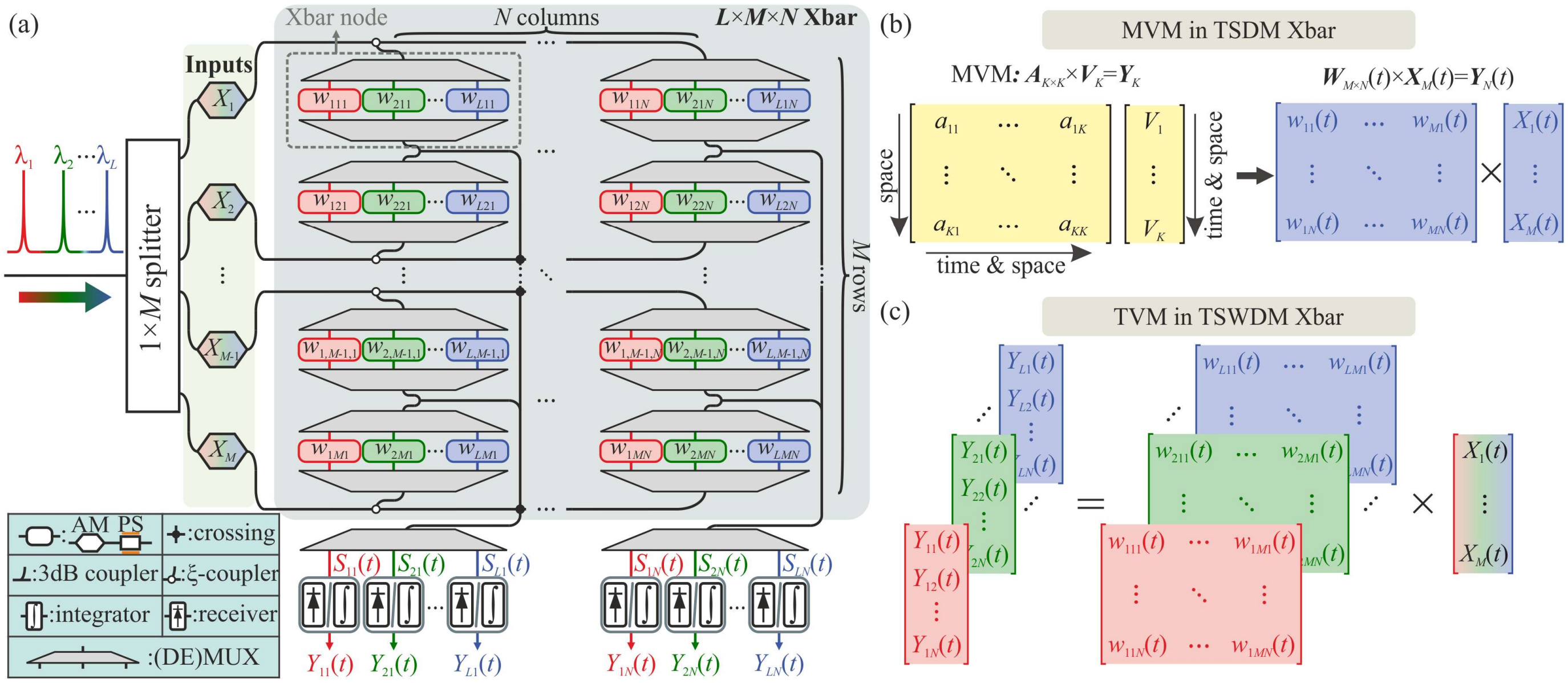

Fig. 2. (a) The $L\times M\times N$ TSWDM Xbar layout, supporting time, space and wavelength division multiplexing. (b) Operation of a single TSDM Xbar for arbitrary MVM computations (c) Operation of the TSWDM Xbar as a TVM engine.

leaving unresolved the computational throughput scaling of the photonic architecture. Combining all three multiplexing techniques within the same PNN architecture can eventually shape a mature path to scale the computational power while supporting a large number of trainable parameters, as has been shown recently in [18],[20],[21]. However, state of the art experimental demonstrations with computational powers exceeding a few hundred TOPS have been only shown via the use of bulk fiber-based devices, as shown in Fig. 1, with on-chip photonic systems still struggling to enable the simultaneous employment of time, space and wavelength multiplexing dimensions.

Transferring these multiplexing concepts on coherent linear photonic circuits can facilitate a latency reduction against state of the art, but this has been almost impossible to realize via SVD-based MZI mesh architectures. The only coherent photonic AI processor engine that managed to combine time as an additional dimension to space has relied on the photonic Xbar architecture [22]-[25]. Multiplexing both time and space dimensions allowed for the demonstration of significant latency reduction against TDM schemes, still, however, enabling only matrix-vector multiplications at a computational power that is constrained by the spatial Xbar circuit dimensions.

In this paper, we present a novel time-space-wavelength division multiplexing (TSWDM) Xbar that can support tensor vector multiply operations in photonic neural networks. Exploiting the potential of the photonic Xbar to perform any MV product at high line-rates [24] with fully restorable fidelity [22], we present a WDM silicon photonic Xbar that employs wavelength-parallelized elementary computational nodes to provide the wavelength dimension on top of the time- and space-multiplexed capabilities of the Xbar architecture. A 4-channel, 2-input silicon photonic Xbar that uses high speed electro-absorption modulators (EAM) as its fan-in and weighting nodes is employed for the proof-of-concept experimental validation of the photonic tensor core operation, revealing accurate tensor representations with an experimental error of <3.9%. The same photonic integrated circuit (PIC) was then utilized for inference operation in the Iris dataset classification tasks at data rates up to 4×60 GBd, supporting a total computational power of 0.96 TOPS on chip and a footprint efficiency of 480 GOPS/mm$^2$. Successful classification of the Iris dataset was then experimentally validated with a high accuracy of 93.3% at data rates of 4×10 up to 4×30 GBd that gradually descends to 83.3% as the line rate increases to 4×60 GBd. Finally, the scalability analysis of the architecture was performed, revealing significant benefits in terms of laser power requirements and achievable bit resolution when a WDM scheme is incorporated in the photonic Xbar.

The rest of the paper is organized as follows: Section I presents the system architecture and the computational workflow of the TSWDM Xbar. Section III presents the experimental demonstration of the 4×2×1 TWDM Xbar, including the description of the photonic integrated circuit and the experimental setup, along with the DC and RF experimental results. Finally, section IV includes the comprehensive discussion on the performance and the scalability potential of the proposed layout.

## II. The Time-Space-Wavelength Multiplexed Xbar

The architecture of the TSWDM Xbar is presented in Fig. 2(a). It is formed by generalizing the coherent photonic Xbar [27] to support simultaneously time, space and wavelength division multiplexing, synergizing the recently demonstrated time-space multiplexed Xbar [24] with the so far only theoretically introduced and analyzed WDM Xbar configuration [28],[29]. It consists of a multiwavelength source, an $L \times M \times N$ Xbar mesh (where $L$, $M$, and $N$ correspond to the number of Xbar's wavelength channels, rows and columns respectively), and a demultiplexing detection site. Firstly, the multiwavelength source provides the $L$ different channels required for the WDM operation, along a common waveguide. The coherent TSWDM Xbar contains an array of $M$ input amplitude modulators (AM) $X_i$, followed by a grid of multiplexed computational nodes arranged in a crossbar layout. Specifically, each node is constructed by interleaving an array of elementary nodes $w_{lmn}$, typically a pair of an AM and a phase modulator (PM), between a demultiplexing and multiplexing stage, where $l$ indicates the respective wavelength channel and $m, n$ the row and column location of the WDM node respectively. All AMs are driven at the same data rate to allow TDM capabilities. Finally, the detection site consists of a demultiplexing stage, to isolate the different wavelengths into different data streams, each containing an optical receiver for electrooptical conversion, and an integrator unit capable of performing temporal accumulations.

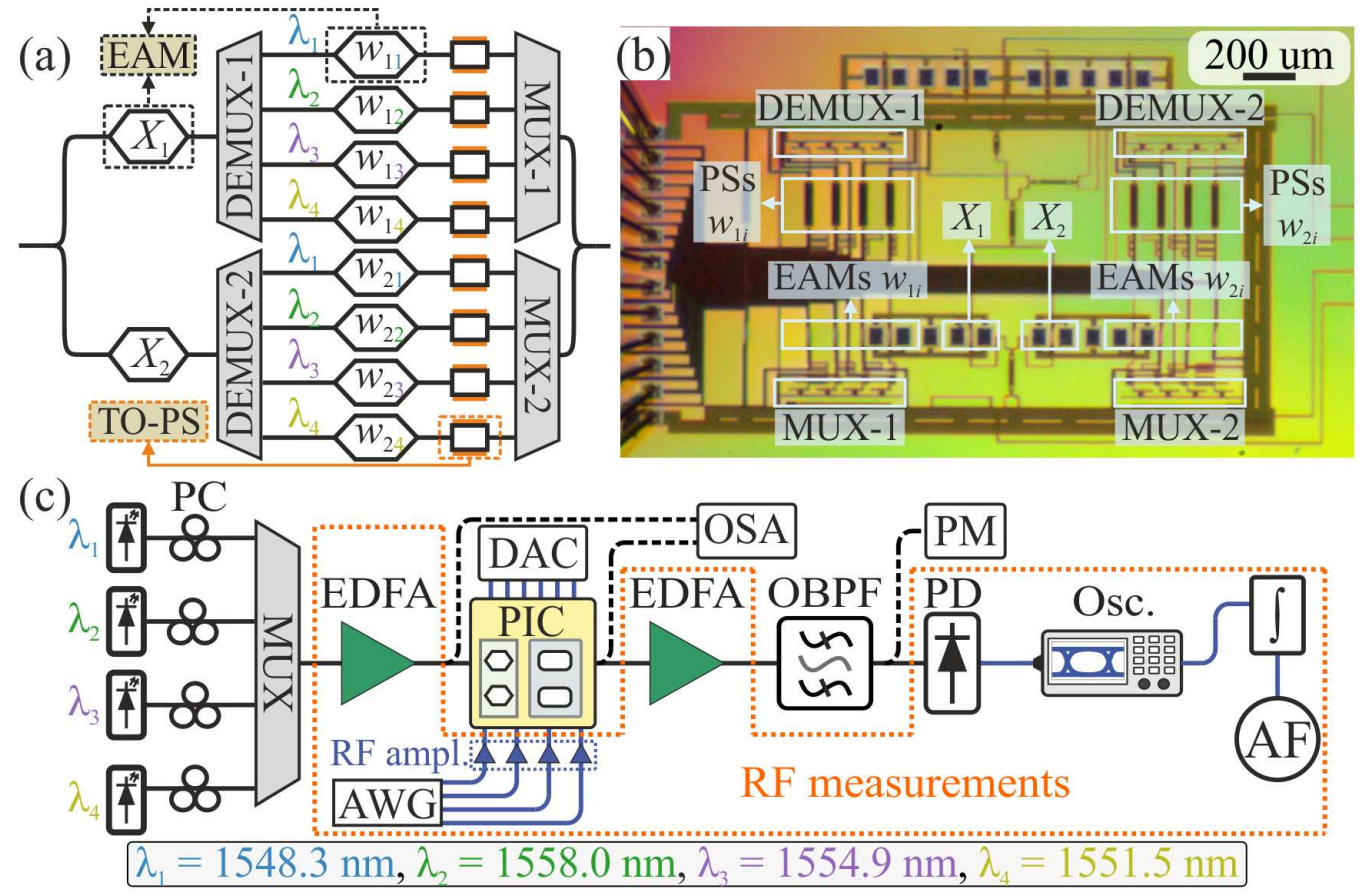

Fig. 3. (a) The 2×1 TWDM Xbar. (b) Fabricated SiPho chip, incorporating high speed EAMs and TO PS (c) Experimental setup for the characterization of the photonic processor.

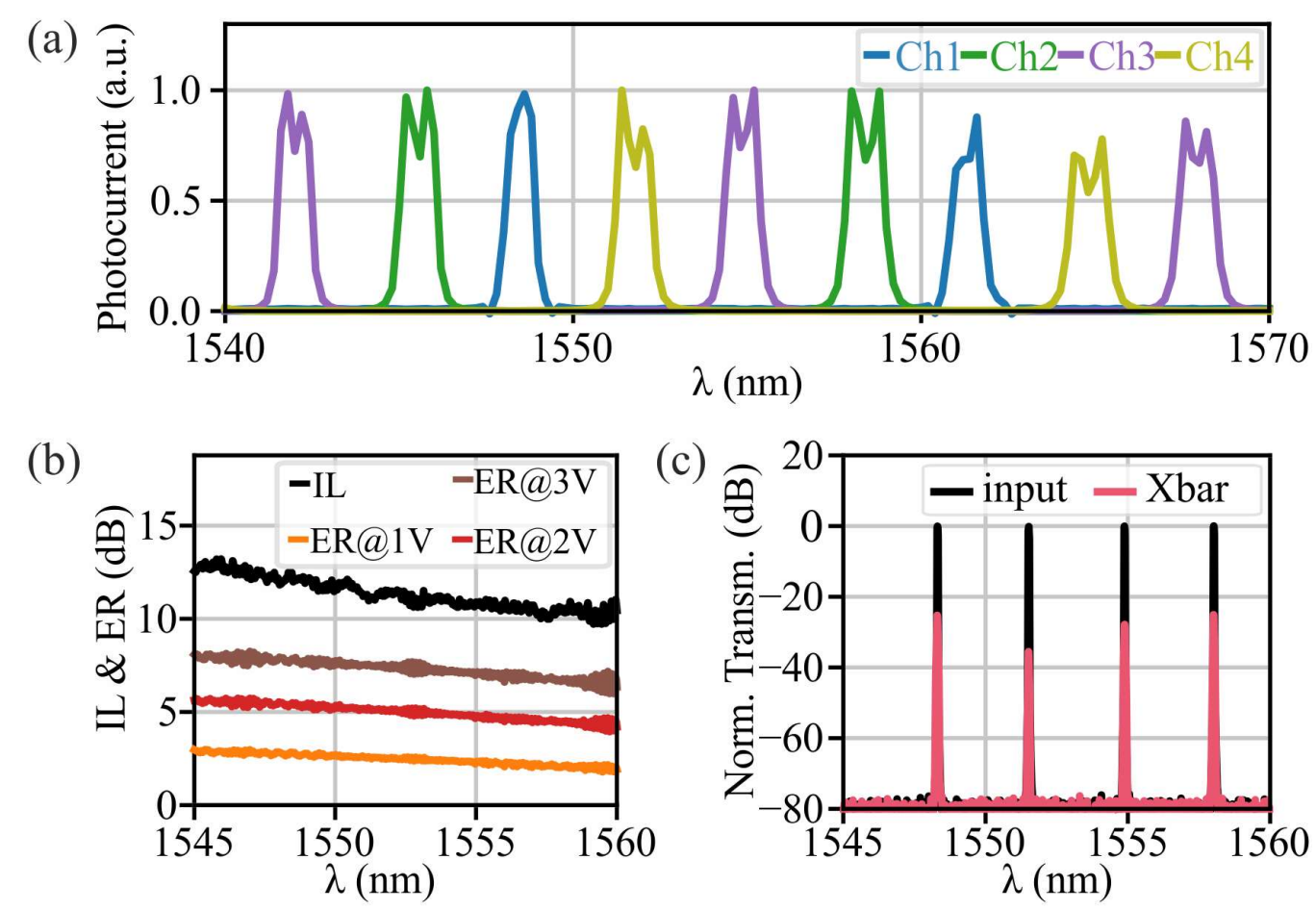

Fig. 4. (a) EAM normalized photocurrents, recorded at the respective weight EAMs, depicting the MUX-DEMUX channel characteristics. (b) IL and ER GeSi EAM. (c) Xbar chip normalized input and chip output transmission spectra.

The operation of the TSWDM Xbar extends the operation of the single-wavelength Xbar-based generic matrix multiplier (GeMM) [24] through the addition of multiple wavelength channels, which provide depth as a third dimension to the matrix layout. In particular, a MVM of the form $\boldsymbol{A}_{K\times K} \times \boldsymbol{V}_K$, where $K \gg M, N$, can be calculated directly at a specific wavelength channel, by simultaneously serializing and distributing the matrix and input vector onto the computational nodes of the Xbar, as shown in Fig. 2(b). The addition of multiple channels in the TSWDM Xbar adds depth in the available computational space, as shown in Fig. 2(c), permitting the computation of any TVM through the photonic hardware.

## III. Experimental Demonstration of the TWDM Xbar

### A. The 4-channel 2-input TSWDM Xbar PIC

For the demonstration of the TSWDM Xbar, we focus on the four-channel, two-input Xbar, presented in Fig. 3(a). The photonic chiplet, fabricated in IMEC's 200 mm platform, incorporated SiGe 50 um, 56 GHz Franz-Keldysh electro-absorption modulators (EAM) as input and weight modulators. To ensure the proper sign imprinting of the weight values, 170 um thermo-optical phase shifters (TO-PS) were included in the Xbar nodes, to ensure constructive interference at each wavelength channel. On-chip, four channel, dual-microring (DE)-MUX stages served as the optical multiplexing stages of the Xbar, providing 3 nm adjacent channel spacing [30]. The MUX response could be thermo-optically tuned externally, leading to the shift of the multiplexed channels and allowing for the alignment of multiple components on a common wavelength grid. The fabricated silicon chip is presented in Fig. 3(b), with the TSWDM Xbar spanning a total area of 2 mm$^2$.

### B. Experimental setup

The setup deployed for the experimental evaluation of the TSWDM Xbar is illustrated in Fig. 3(c). A 4-channel tunable laser source (IDPhotonics CoBrite DX4) generated four light beams, with wavelengths $\lambda_1$=1548.3 nm, $\lambda_2$=1558 nm, $\lambda_3$=1554.9 nm and $\lambda_4$=1551.5 nm, that were coupled into a single fiber through a packaged MUX stage and injected into the PIC. The optical input and output signals were monitored either by an optical spectrum analyzer (OSA) or by an optical power-meter (PM). The integrated EAMs' bias voltage and the current of the thermo-optical elements were controlled through a multi-channel digital to analog converter (DAC) panel, whereas source measure units (SMUs) were deployed to monitor the photocurrent of the EAMs.

For the NN-task experiments, a 4-channel Keysight M8194a arbitrary waveform generator (AWG), equipped with four SHF S804b electrical amplifiers, was incorporated into the setup to drive the RF components of the photonic Xbar. The modulated output was captured by a 70 GHz photodiode (PD) and recorded by a Keysight N1046a sampling scope, whereas the integrator and activation function (AF) units were implemented offline. A series of polarization controllers (PC), optical band-pass filters (OBPF) and erbium doped fiber amplifiers (EDFA) were included in the setup to ensure the signal integrity of the optical link.

### C. DC characterization –Tensor-Vector multiplier

The wavelength (DE)MUX capabilities of the TSWDM Xbar PIC were evaluated by aligning the integrated (DE-)MUX structures via monitoring of the photocurrent of the weight EAMs. Since there is a one-to-one correspondence between the value of the EAM photocurrent and the incident optical power to the under-measurement EAM, the channel profile of the integrated DEMUX stages can be constructed. For the alignment of the MUX stages, the same procedure was applied by injecting the tunable CW beam from the output port of the PIC. An indicative set of normalized photocurrent measurements versus wavelength for an aligned MUX stage is presented in Fig. 4(a), confirming the spectral characteristics of the integrated multiplexers.

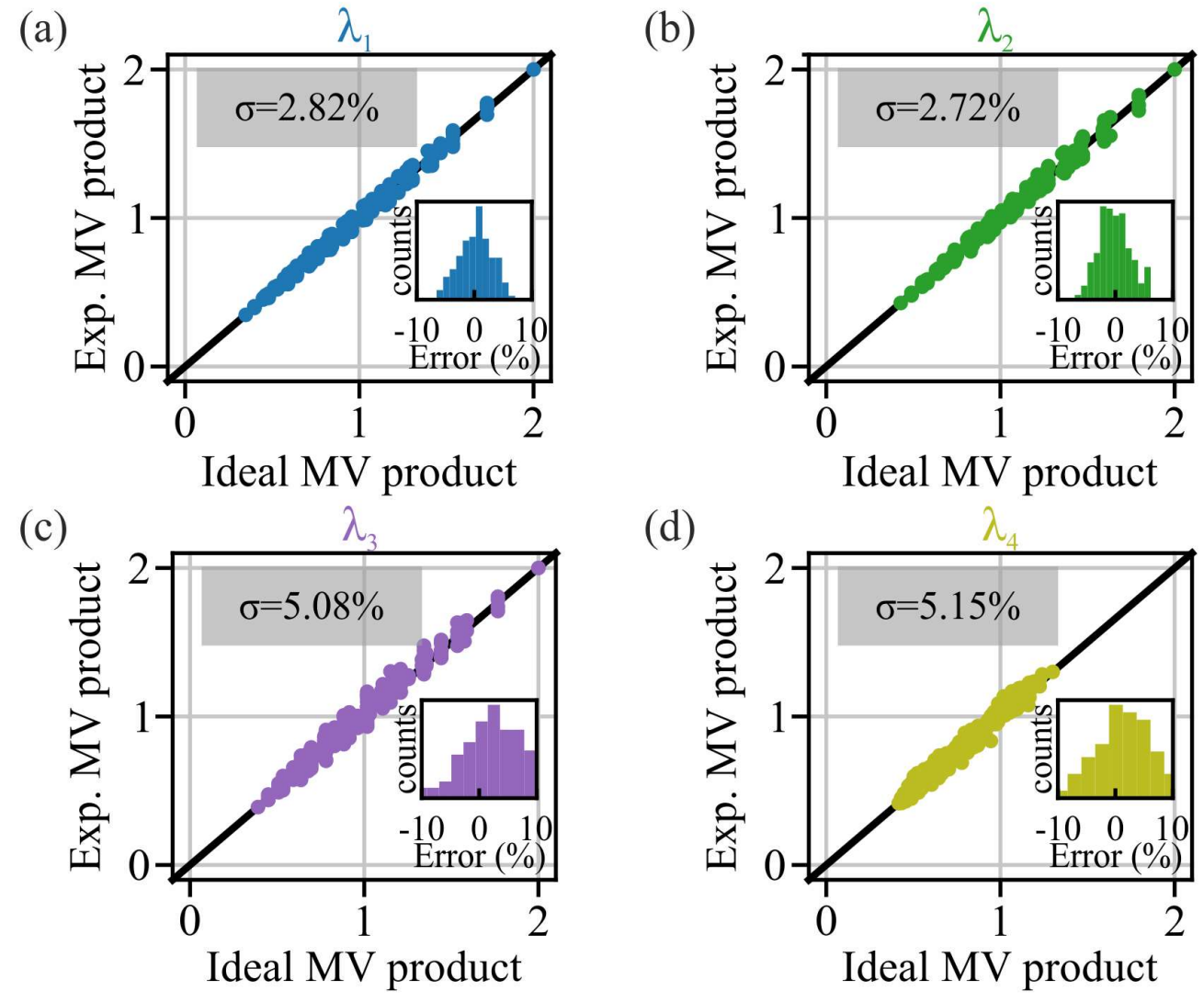

Fig. 5. MVM operations under injection of four different channels, depicting an average error of 3.9%.

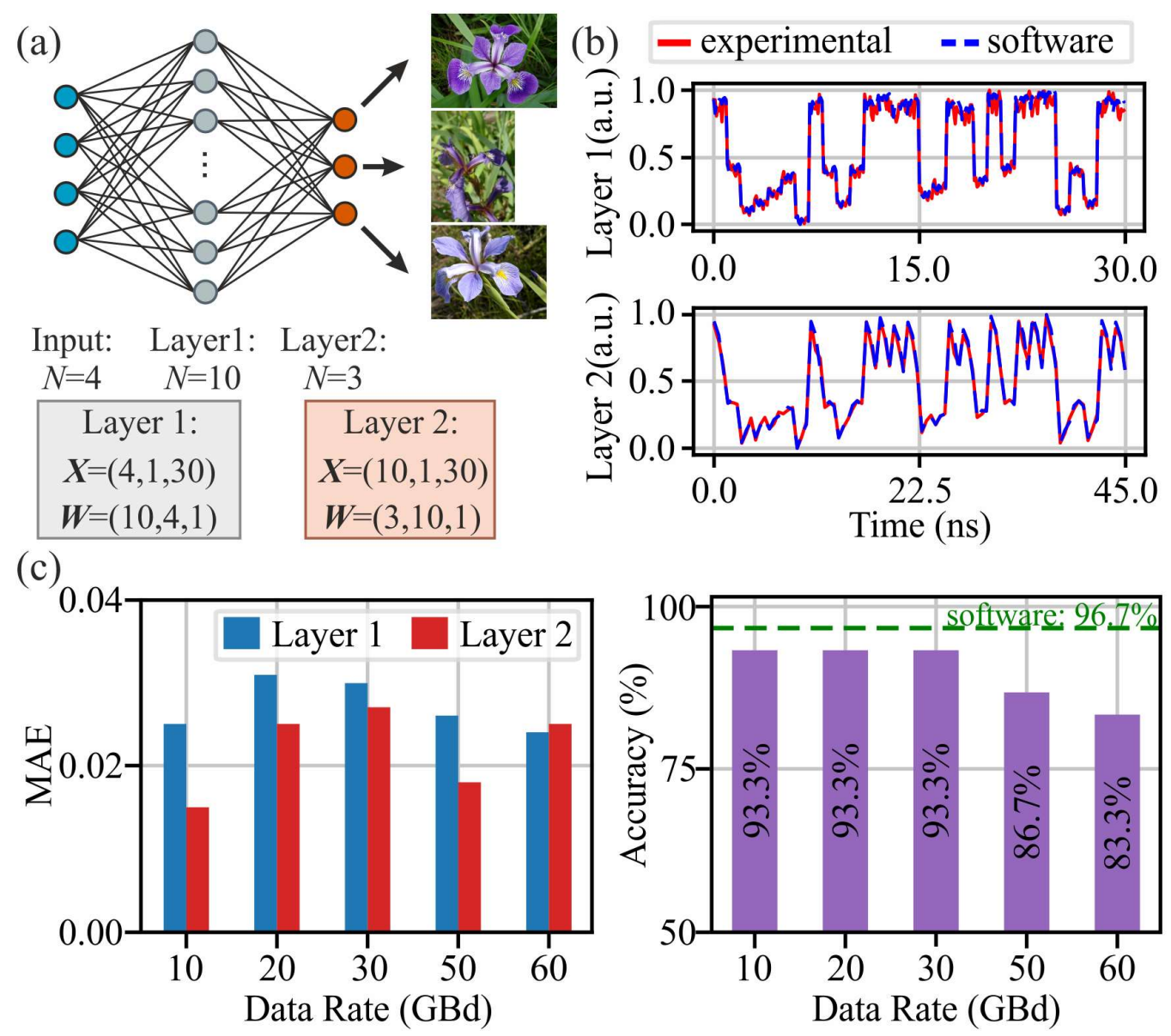

Fig. 6. (a) Fully connected NN, used for the IRIS classification task. (b) Output traces of the different NN's layers at a compute rate of 10 GBd. (c) MAE per layer and total NN accuracies for 10, 20, 30, 50 and 60 GBd

The performance of the EAMs in the TSWDM Xbar was evaluated through a series of DC measurements, revealing the insertion loss (IL) and static extinction ratio (ER) of the modulators. As presented in Fig. 4(b), the IL of a single EAM lies in the range of 11 dB for wavelengths of 1545 up to 1560 nm. The static ER at 1550 nm reached approximately 2.5 and 7.5 dB for a reverse bias voltage of 1 and 3 V respectively. The EAM IL has been significantly higher compared to the typical IL values of 6-7dB that are experienced by respective single-wavelength silicon photonic Xbar PICs [25],[26], probably originating from a slightly differentiated fabrication process, still allowing, however, for the proof-of-concept demonstration of the TSWDM Xbar. The optical signals were injected into the photonic chip with balanced powers, with the normalized transmission of the input and output spectra presented in Fig. 4(c). Furthermore, the normalized Xbar output spectrum indicates a balanced performance, excluding channel $\lambda_4$, which presents significantly higher losses due to the degraded performance of the $w_{24}$ EAM.

The operation of the TWDM Xbar as a Tensor–Vector multiplier was demonstrated through a series of DC measurements, summarized in Fig. 5. By incorporating a single wavelength $\lambda_r$ into the TWDM Xbar, both the input modulators $X_{1,2}$ and the respective weight modulators $w_{1r}, w_{2r}$ were biased with voltages in the set {0,-1,-2,-3} V, with $r$=1,2,3,4 denoting the respective wavelength channel, leading to 256 distinct dot products per wavelength channel. The experimentally obtained products, computed after calibrating the TSWDM crossbar to restore the non-linearity and non-uniform IL of the modulators, were compared with the ideally expected products, computed for perfect modulators with non-infinite ER [22]. The clear linear correlation between the experimental and ideal products is evident from our measurements, further supported by small average error values $\sigma$=3.9%, calculated as $\sigma = (P_{exp} - P_{ideal})/P_{ideal} \cdot 100$ %. A smaller range of experimentally obtained products could be obtained for channel $\lambda_4$ of the TSWDM Xbar, predominantly due to the degraded performance of the $w_{24}$ EAM.

### *D. The TSWDM Xbar in PNNs*

The performance of the TSWDM Xbar in NN tasks was evaluated through a fully connected NN, trained for the classification of the IRIS dataset [23]. The network topology, illustrated in Fig. 6(a), includes 4 inputs with its first layer comprising 10 neurons and its second layer consisting of 3 output neurons. Emulating the inference procedure through the photonic hardware, 30 different batches were processed through the TSWDM Xbar at different data-rates from 10 up to 60 GBd. Both the input and weight modulators were driven with RF signals having a peak-to-peak voltage between 2-3Vpp . The temporal integrator and sigmoid AF units were implemented offline. Indicatively, we present the NN output traces for a data rate of 10 GBd in Fig. 6(b), with the the experimental trace closely following the respective traces obtained by software. The mean absolute error (MAE) of each layer and the experimental accuracy of the NN were also measured for all different data rates and are summarized in Fig. 6(c). As illustrated, the TSWDM Xbar presents an overall stable performance for all data rates with a low MAE value that is centered around a mean value of 1.8% for both layers. The experimental accuracy was found to be constantly at 93.3% for all data rates up to 30 GBd, being only slightly lower than the 96.7% accuracy obtained by software and showing a minor degradation reaching 86.7 and 83.3% accuracy values as the data rate increases to 50 and 60 GBd, respectively.

## IV. Discussion

We have presented the first coherent hyper-multiplexed photonic processor, capable of performing arbitrary MVM and TVM computations, for data rates up to 60 GBd and computational throughput of 0.96 TOPS, by incorporating time, space and wavelength division multiplexing. Even though spatial and wavelength dimensions hold an equivalent role in the proposed layout, they implicitly dictate the scalability potential of the architecture.

Table I. Bit resolution analysis parameters.

| | | | |
|---|---|---|---|
| $R_{PD}$ | 0.8 A/W | $IL_{EAM}$ | 5 dB [33] |
| $RIN$ | -150 dB/Hz [31] | $IL_{\xi}$ | 0.2 dB [34] |
| $\Omega$ | 60 GHz | $IL_X$ | 0.02 dB [35] |
| $q$ | 1.6 $10^{-19}$ C | $IL_{MMI}$ | 0.06 dB [36] |
| $ER$ | 8 dB [33] | $IL_{MUX}$ | 0.2 dB [37] |
| $i_{ref}$ | 1.5 $10^{-11}$ dB/$\sqrt{\text{Hz}}$ | | |

In more detail, the computational throughput of an $N \times N$ SDM Xbar, will be equal to:

$$OPS = 2N^2\,\Omega \quad (1)$$

where $\Omega$ is the compute rate of the Xbar's modulators and the factor of 2 stands for converting the MAC/sec of $\Omega$ into OPS, since 1 MAC/s=2 OPS. Similarly, an $L \times N \times N$ WDM Xbar will present a computational throughput of:

$$OPS = 2L\,N^2\Omega \quad (2)$$

This means that the $N \times N$ SDM Xbar can have the same computational power with an $L$-channel WDM Xbar only if it comprises a number of rows and columns that is increased by a factor of $\sqrt{L}$ compared to the number of rows and columns of the WDM Xbar version. Despite the fact that spatial and spectral degrees of freedom are interchangeable in terms of computational power, the two scalability approaches differ when scaling the photonic circuitry in a, performance-wise, sustainable route.

A suitable metric that reflects the scalability perks of the WDM scheme is the achievable bit resolution $n_{bit}$ at the output of the photonic Xbar, after optoelectrical conversion. The achievable bit resolution at the output of the Xbar [25] will be equal to:

$$n_{bit} = \log_2\left(\frac{OMA}{\sqrt{12}\sigma_{total}} + 1\right) = \log_2\left(\frac{2P_{avg}}{\sqrt{12}\sigma_{total}}\frac{ER-1}{ER+1} + 1\right) \quad (3)$$

where $OMA$ is equal to the optical modulation amplitude of the analog output of the photonic Xbar, $P_{avg}$ is the average optical power at the output of the Xbar, $\sigma_{total}$ is the total noise magnitude present at the captured signal and $ER$ is the extinction ratio of the modulators. Assuming that all noise sources can be considered as Gaussian, and attributed to the relative intensity noise (RIN) of the laser, the shot noise of the PD and the thermal noise of the transimpedance amplifier (TIA), the total noise in the captured signals will be:

$$\sigma_{total}^2 = \sigma_{RIN}^2 + \sigma_{PD}^2 + \sigma_{TIA}^2 = \left(\left(R_{PD}P_{avg}\right)^2 RIN + 2q\left(R_{PD}\cdot P_{avg}\right) + i_{ref}^2\right)\Omega \quad (4)$$

where $R_{PD}$ is the responsivity of the PD, $P_{avg}$ is the average optical power reaching the PD, $RIN$ is the relative intensity noise density of the laser, $q$ is the electron charge and $i_{ref}$ is the current noise level of the TIA. For a specific choice of components, the only parameter that depends on the physical dimensions of the photonic accelerator is the optical power reaching the PD, depending directly on the IL of the different Xbar architectures and the provided laser power.

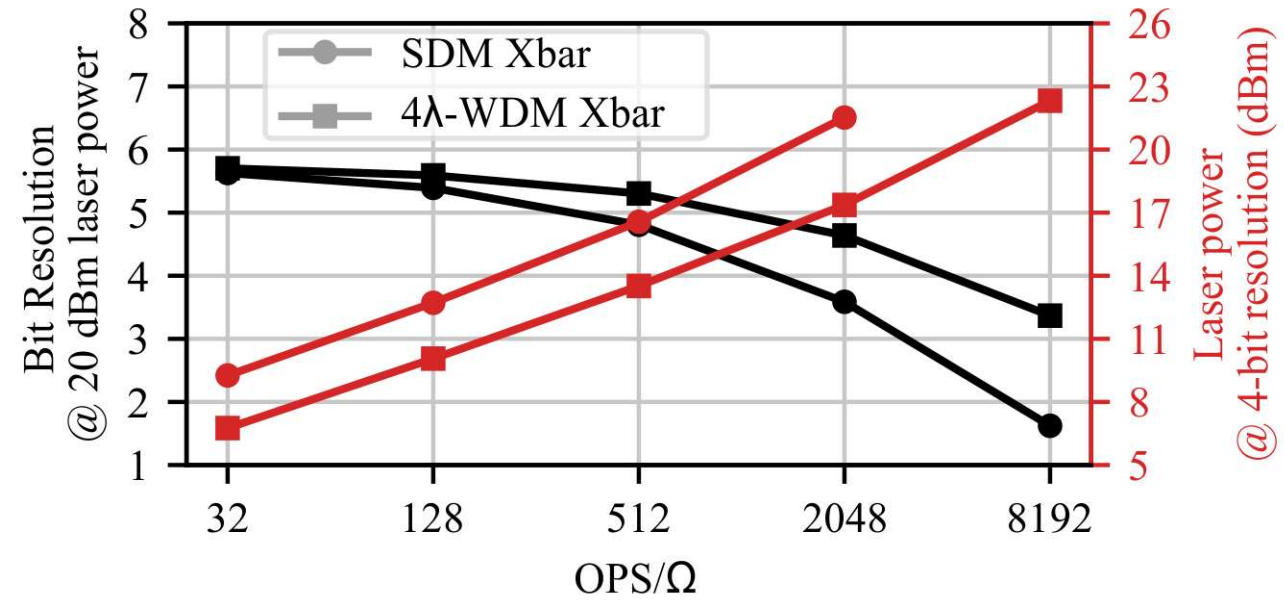

Fig. 7: Bit resolution and single laser power for optimal performance of a single wavelength Xbar and a $4\lambda$ WDM Xbar, as a function of the normalized computational throughput.

Specifically, the IL of the $N \times N$ SDM Xbar [27] will be equal to:

$$IL_{SDM} = 2IL_{EAM} + IL_{\xi} + \left(\frac{N}{2} - 1\right)IL_X + (2\log_2 N)IL_{MMI} - 10\log_{10}\xi_1^2 \quad (5)$$

where $IL_{EAM}, IL_{\xi}, IL_X, IL_{MMI}$ are the IL of the EAM, $\xi$-coupler, crossing and 3-dB coupler, whereas $\xi_1$ is the characteristic splitting ratio accounting for a power balanced Xbar layout, where all crossbar columns provide the same output power level [22],[27].

Moreover, the IL of a single channel of the $L \times N \times N$ WDM Xbar is:

$$IL_{WDM} = IL_{SDM} + 4IL_{MUX} \quad (6)$$

where $IL_{MUX}$ is the insertion loss of the integrated (DE)MUX stages. It was assumed that $L$ different single laser sources are multiplexed into a common route at the input port of the WDM Xbar. Based on the above relationships, and considering state of the art components, the achievable bit resolution for different single laser powers can be calculated for the two flavors of the Xbar architecture. The parameters used in the analysis are summarized in Table I.

From the previous equations, it is evident that incorporating a WDM scheme to the photonic Xbar will yield greater losses than the SDM approach of the same radius $N$. However, a more appropriate comparison between the different Xbar flavors must be performed with respect to the computational throughput, which reflects the performance of the two architectures. The results of our analysis are summarized in Fig. 7, for the case of an $N \times N$ SDM Xbar and a $4\lambda \times N' \times N'$ WDM Xbar. Firstly, by setting the single laser power to 20 dBm [38], the WDM Xbar presents improved bit resolution compared to the SDM Xbar for the same computational throughput (normalized to the compute rate $\Omega$), with a difference up to 1-bit observed for scaled up architectures. This is due to the fact that the WDM Xbar exhibited up to 10 dB IL reduction compared to the SDM case, primarily because the increased Xbar dimensions require more crossings. Moreover, scaled up WDM layouts can provide moderate bit resolutions for PNNs, equal to 4-bit resolution for 1024 OPS/Ω and 3-bit resolution at 4096 OPS/Ω, projected to 122.9 TOPS and 491.5 TOPS at 60 GBd compute rates respectively. Finally, an additional advantage regarding the potential feasibility stems from our analysis, since the WDM layouts require lower laser power for a specific targeted performance, compared to the SMD Xbar, as depicted in Fig. 7. In particular, the power provided by a single laser for achieving 4-bit resolution at a specific computational throughput improves by a factor of approximately 3 dB when incorporating a WDM scheme in the SDM Xbar.

## V. Conclusion

We have presented a novel time-space wavelength division multiplexing TVM engine for photonic neural networks. The proposed scheme was verified through the experimental study of a 4×2×1 integrated TSWDM Xbar, which demonstrated an average error of 3.9% in TVM computations and experimental accuracy greater than 83.3% for the IRIS classification task for data rates up to 60 GBd. The scalability potential of the architecture was verified through a rigorous analysis of the achievable bit resolution in the photonic TSWDM Xbar, revealing the prospect of achieving 491.5 TOPS computational throughput in scaled-up layouts.